# Ion beam analysis and big data:
# How data science can support
# next-generation instrumentation

Tiago F. Silva[1], Cleber L. Rodrigues, Manfredo H. Tabacniks,
Hugo D. C. Pereira, Thiago B. Saramela, Renato O. Guimarães

*Physics Institute of the University of São Paulo, São Paulo, Brazil.*

**Abstract:** With a growing demand for accurate ion beam analysis on a large number of samples, it becomes an issue of how to ensure the quality standards and consistency over hundreds or thousands of samples. In this sense, a virtual assistant that checks the data quality, emitting certificates of quality, is highly desired. Even the processing of a massive number of spectra is a problem regarding the consistency of the analysis. In this work, we report the design and first results of a virtual layer under implementation in our laboratory. It consists of a series of systems running in the cloud that perform the mentioned tasks and serves as a virtual assistant for member staff and users. We aim to bring the concept of the Internet of Things and artificial intelligence closer to the laboratory to support a new generation of instrumentation.

# Introduction

Since the early ages of ion beam analysis techniques, state of the art for detectors, electronics, and associated equipment (like rastering systems and automated sample holders) considerably changed over the decades, enabling a substantial increase in the amount of stored data recorded during the experiments [1]. Currently, at the Laboratory of Material Analysis with Ion Beams of the University of São Paulo (LAMFI-USP), not just spectral data from each detector are recorded. Still, a computer-assisted system monitors some main parameters of the particle accelerator during each experiment, and all this information is stored into a database accessible through a website interface. This database also includes environmental information like laboratory temperature and humidity.

Besides that, the commissioning of the new external beam infrastructure with capabilities for scanning large areas [2] introduced a new primary source of data in the laboratory. It produces a large amount of information for each map that is virtually impossible to be processed by humans. Data includes photos of each analyzed spot prior and after irradiation, spectra of visible luminescence, surface coordinates in the XYZ-axis, two x-rays detectors, one gamma detector, the charge normalization information, among others.

---

1  Corresponding author: tfsilva@if.usp.br

In this work, we report the design and partial construction of a software layer in the laboratory infrastructure to pre-process the experimental data, aiming to produce, as quick as possible, useful feedback on the experiment itself, and on the conditions in which the experiment took place.

The system starts with data acquisition software that also does data management, header preparation, and data conversions. Currently, a Data Quality Assurance System (DQAS) is under implementation. The system checks accelerator conditions during every experiment, checks the quality of the spectra, and also checks if there is some correspondence to the header information within the spectra files. If any step presents failure, the system advises the accelerator staff to solve the issue. Additionally, sanity checks are made based on the recorded data, automatically producing reports on the health status of the accelerator and enabling maintenance schedules. All this infrastructure is oriented to provide the user with promptly ready files for a self-consistent analysis through MultiSIMNRA [3].

Regarding spectra processing, the system in the cloud counts with a set of previously trained artificial neural networks to process scattering analysis of some typical and uncomplicated cases. These networks can predict layer thicknesses, composition, and roughnesses according to training sets generated by simulations with SIMNRA [4]. They are automatically triggered when the data is saved in the cloud storage and configured by information inserted by the user when filling the beam time form on our website. Even though ANN is proved as useful in processing nuclear scattering spectra for material analysis [5–8], this is the first report of its implementation in production as part of the analysis routine.

In the particular case of large-area maps, we use machine learning techniques to detect similarities between the pixels automatically. Summing the similar pixels as determined by a clustering algorithm increases statistics and improves detection limits. It also enables the identification of patterns in the recorded spectra, enabling the discovery of the correlation between elements along the mapped area and revealing nuances hardly observed by the naked eye [9]. A similar implementation can be found elsewhere [10].

Our findings support that this approach enhances the user experience and the convenience of data analysis. It brings the concept of the Internet of Things (IoT) closer to the laboratory as a virtual assistant of analysis, opening a wide range of possibilities for new generation instrumentation.

# Methods

Aiming at a better efficiency in data management and processing, we designed a cloud system to act as a virtual assistant. This system is composed of three subsystems: one for data acquisition and management, another for quality assurance and performance checks, and the last for spectra processing of more straightforward cases.

This architecture targets the expansion of the data processing capabilities supporting a massive throughput keeping the quality standards regarding traceability and consistency of analysis. As an additional feature, the system can do online processing of the data emiting alerts to the operator if an unusual situation takes place.

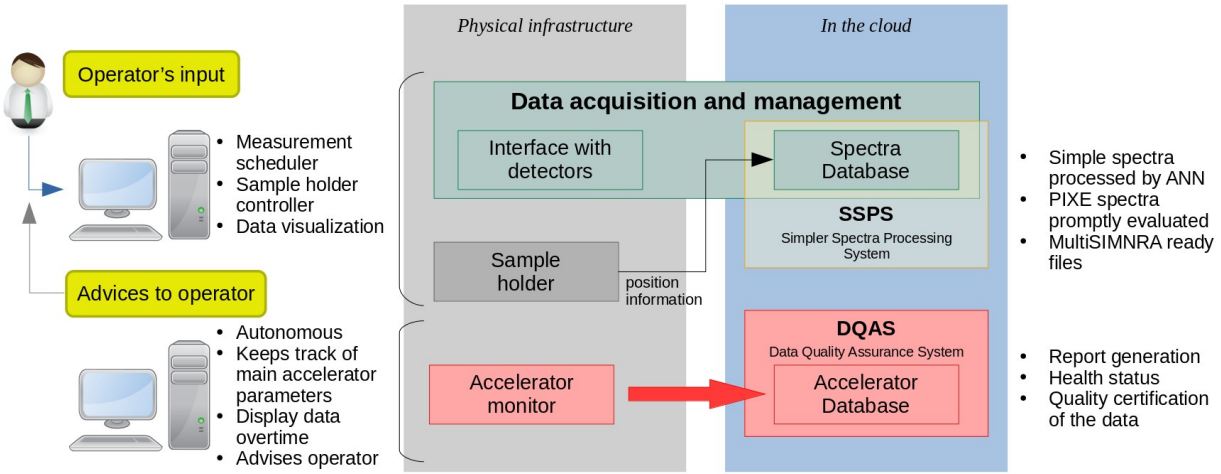

Figure 1 – Schematic of the cloud infrastructure and its interaction with physical devices. SSPS stands for Simpler Spectra Processing System, and DQAS stands for Data Quality Assurance System.

## Data acquisition and management

It is in an advanced level of development a custom-made data acquisition software. It supports a scalable architecture of digitizers (currently tested with two 8 channels each CAEN model N6725) working in high count rates. Automatic control of transfer rate of data that takes into account the detector counting rates of each channel and the memory buffer size of the digitizer provides comfortable use of the software, with no overload of the computer CPU, and with adequate refreshing time of the displayed spectra. The software also records the data in list mode enabling histogram reconstructions.

The acquisition software also communicates with a picoammeter (CAENels model AMR549, with 4 channels, 24-bit A/D conversion, up to 26 kHz sampling frequency, and 300 fA resolution), enabling the record of the beam current incident on the target overtime of the experiment. Stop criteria available include total acquisition time and total accumulated charge (calculated by numerical integration of the current signal).

The system performs well with surface barrier detectors (using preamplifiers CAEN model A1422B090F2, 90mV/MeV gain, Cdet < 200pF), and implementation for x-rays detectors is under development. In the present architecture, up to 16 detectors can be monitored, with pulse height analysis by trapezoidal fit method or pulse shape discrimination by charge integration gates widths method, both as provided by firmware distributed by the digitizer manufacturer.

An additional and essential feature is a TCP server either for remote control of the acquisition or for the remote communication with different subsystems, like sample holders and acquisition scheduler. A schematic of the software architecture is depicted in fig. 2.

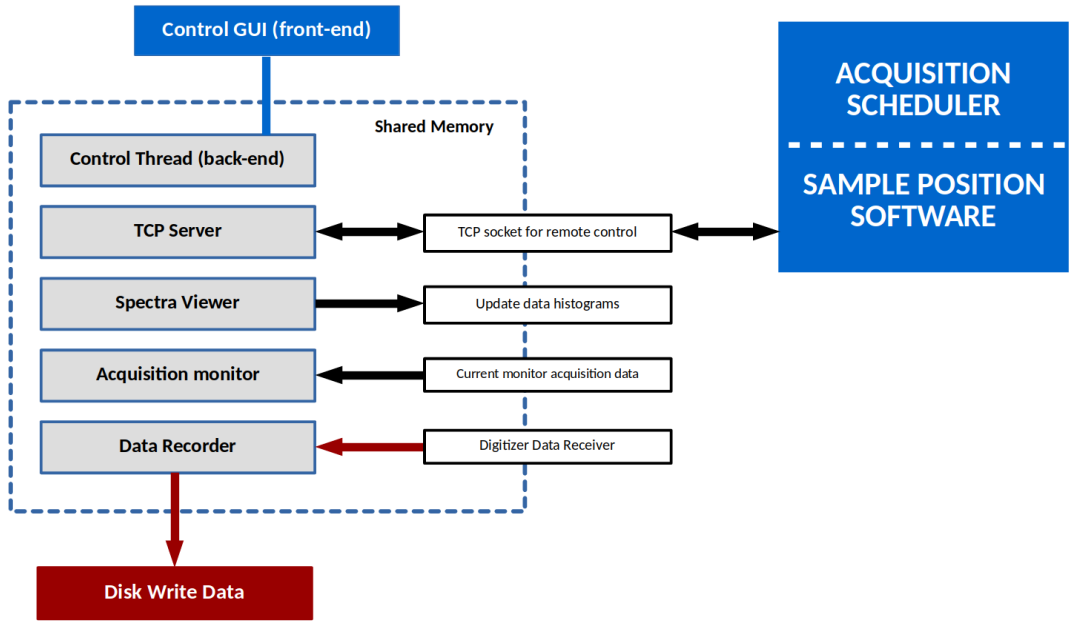

Figure 2 – Schematic of the internal structure of the custom made software for data acquisition for the LAMFI-USP facility. Control of rate of data transfer and TCP server for external communication are the main features of the software.

This software also features tools oriented to data management (serial labeling, naming, user history, etc.) and to provide data in formats compatible with the most widely used software (SIMNRA, MultiSIMNRA, NDF, WinAXIL, RUMP, etc.). It also offers to the user with promptly ready files for a self-consistent analysis through MultiSIMNRA [3].

## Data Quality Assurance System (DQAS)

The DQAS is a system that is permanently keeping track of the accelerator conditions. It can process the data and be a source of two types of information: on the conditions of the accelerator overtime of each experiment, and on the health status of the accelerator in a longer time window comparison (weekly or monthly).

The system checks accelerator conditions during every experiment, checks the quality of the spectra, and also checks if there is a correspondence between the header information within the spectra files with the accelerator condition. If any step presents failure, the accelerator staff is advised to solve the issue. Many sorts of solutions can be adopted, ranging from merely fixing the file header (human error), up to the removal of data during specific time windows. Since data in list mode is available for both detectors and beam current (see the previous section), this enables the use of a tool to histogram the data again, removing the defective part, thus minimizing the necessity to repeat the measurement.

Additionally, sanity checks are made based on the recorded data, and reports are generated automatically and periodically on the health status of the accelerator, enabling better maintenance schedules. By comparing the performance indicators over 2 to 6-months time windows, it is possible to detect machine wearing with much more sensitivity than operators' expertise.

## Simple Spectra Processing System (SSPS)

The Simple Spectra Processing System (SPSS) targets the automatic spectra analysis of chronic cases. Some samples have multiple technological interests, and because of that, they are produced in large amounts by many different groups. These samples are analyzed either by nuclear scattering techniques or by PIXE to assess the same characteristics. Such cases demand human resources for spectra interpretation and analysis that could be done in an automated fashion.

The design of SPSS includes the storage of previously trained ANNs and the use of them to process IBA spectra according to the designation. As the generalization capability of each network is limited to a single case, growing the number of trained networks expands the SPSS utility.

For the nuclear scattering cases, the training sets for all ANNs are generated by simulation with SIMNRA with noise added. Thus, one expects that the predictions are consistent with analysis using SIMNRA. The first application results of SPSS were published [11], and comparisons of its performances against human evaluation and evaluation in batch mode are the topic of a different contribution to this conference.

For PIXE analysis, the lack of simulation software avoids the generation of training sets. Currently, it is under intense work in our group the development of simulation software for PIXE that enable PIXE analysis by the SPSS. In fact, few software can perform some PIXE calculation [12,13] in a fashion that does not fit our needs entirely, either because they do not provide the spectra as output, or are not widely available or discontinued.

The system is automatically triggered when the data is saved in the cloud storage and configured by information inserted by the user when filling the beam time form on our website. Even though ANN is proved as useful in processing nuclear scattering spectra for material analysis [6], this is the first report of its implementation in production as part of the analysis routine.

## Machine learning for mappings

Machine learning tools have been adopted to process data of the large-area mapping device [9]. The designed cascade of algorithms can find similar regions (within a certain level of statistical significance) and sum the spectra of each pixel included in this region. It is useful by one side since this procedure improves the detection limit of the technique. On another side, this feature is only available in an average assessment of the concentration of the elements all over particular regions.

The algorithms are also useful in finding correlations over the spatial distribution of elements and detect different contributions to the spectra coming from different components that form the object under analysis. It is an essential outcome since the components can correlate with different material phases, separating contributions from different pigments, paintings, glaze, tissue, layers, among other possibilities, depending on the context of analysis.

A detailed description of the procedure adopted is reported elsewhere [9].

# Results and discussion

Here we present some results of the DQAS. The automatic system to generate monthly reports on the accelerator status is able to process de database searching for unusual situations. Fig. 3 shows the results of the spark counting algorithm, which counts the average number of sparks per working day over the whole dataset. It is possible to observe a systematic increase in the number of sparks during the year of 2019. It was related to an $SF_6$ leakage problem, that reduced the electric insulation. With the problem solved, the number of sparks returned to normal, as indicated in the last bar of the plot.

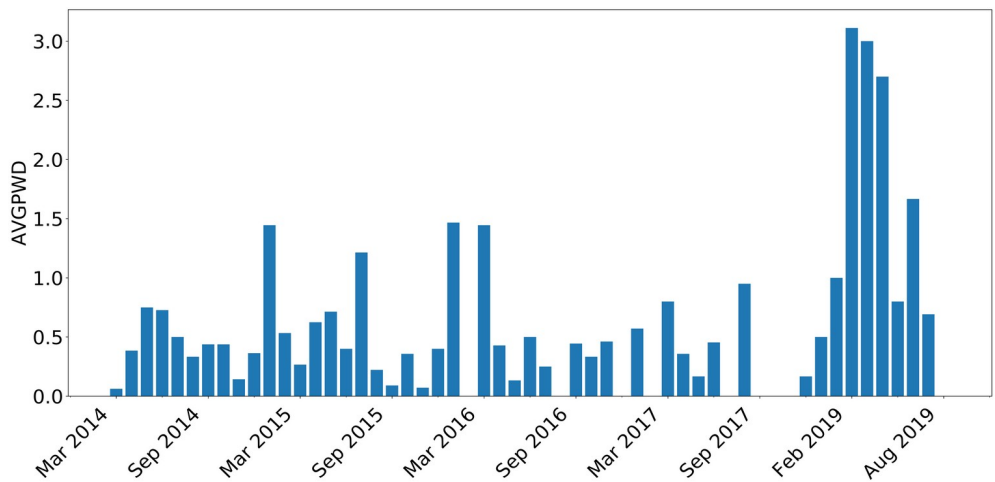

Figure 3 – Average number of sparks in the accelerator terminal per working day (AVGPWD). A systematic increase of sparks during the year of 2019 was related to a leakage problem of the $SF_6$ insulating gas.

Fig. 2 shows a histogram of the ratio for the measured electrical currents that flow through the two columns of resistors that connect the terminal to the ground potential. The expected value is unity, but the residue accumulation from the $SF_6$ decomposition induces change on this value. This accumulation occurs mostly on top of the column on the side of the Van der Graaf generator. Two consecutive months are presented (black and red) for comparison. The red one shows a small tendency for higher values, indicating a possible increment on the deposits together with the $SF_6$ decomposition.

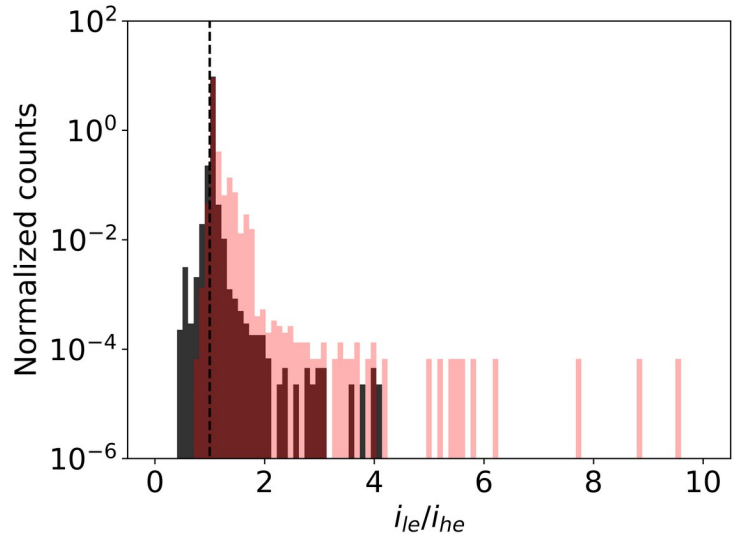

Figure 4 – Histograms of the ratio of the measured electrical currents that flow thru the two columns of resistors that connects the high voltage terminal to the ground potential.

The SPSS has been already utilized to process a massive number of RBS spectra. A study evaluating the performance of SPSS and the use of Artificial Neural Networks to process IBA spectra is presented in a different contribution to this proceeding. Its performance was compared against the human evaluation, and the use of a classical algorithm for optimization running in batch mode. Our findings revealed ANN (thus SPSS too) as more accurate when compared to the human evaluation, and is more efficient (in terms of shorter computing time) when compared with the optimization in batch mode for the case of a massive number of spectra (hundreds or thousands of spectra). The accuracy of SPSS and the batch mode are comparable.

The routines and the cascade of algorithms applied to process PIXE data on mappings are presented in [9].

# Conclusions

We reported the design of a virtual system to process the data produced in an IBA laboratory. The system is designed to be an assistant to the member staff and users.

On its stage of development, DQAS proved to be efficient in detecting abnormal conditions of operation of the accelerator. The spectra quality check and the automatic generation of reports are under implementation.

The SPSS has been used to process a massive number of spectra by employing ANNs. Its performance and efficiency evaluations are reported in a separate contribution to this proceeding.

As a perspective of work, we aim the systems working together autonomously. The evaluation of PIXE measurements by SPSS is also in our plans.

# Acknowledgments

The authors acknowledge the financial supported of University of São Paulo.